\documentclass[journal]{IEEEtran}

\usepackage[utf8]{inputenc}

\usepackage{graphicx}
\usepackage{subfig}
\usepackage[dvipsnames]{xcolor}
\usepackage[markup=underlined]{changes}
\definechangesauthor[color=Blue]{D}
\definechangesauthor[color=Red]{A}
\definechangesauthor[color=Green]{M}

\usepackage{etoolbox}
\newtoggle{draftversion}
\togglefalse{draftversion}

\iftoggle{draftversion}{
	\newcommand{\note}[1]{\textcolor{red}{\sc[NOTE: #1]}}
	\newcommand{\ggx}[1]{\textcolor{CadetBlue}{\sc[L: #1]}}
	\newcommand{\mb}[1]{\textcolor{cyan}{[M: #1]}}
}{
	\newcommand{\note}[1]{}
	\newcommand{\ggx}[1]{}
	\newcommand{\mb}[1]{}
    \renewcommand{\added}[2][]{#2}
    \renewcommand{\deleted}[2][]{}

}

\usepackage{times}  
\usepackage[colorlinks=true,linkcolor=black,anchorcolor=black,citecolor=black,filecolor=black,menucolor=black,runcolor=black,urlcolor=black]{hyperref}

\begin{document}
\title{Distributed Function Chaining with Anycast Routing \\
{\LARGE \textit{Technical Report}}}
\author{\IEEEauthorblockN{Adrien Wion\IEEEauthorrefmark{1}\IEEEauthorrefmark{2},
Mathieu Bouet\IEEEauthorrefmark{1}, Luigi Iannone\IEEEauthorrefmark{2} and
Vania Conan\IEEEauthorrefmark{1}}
\IEEEauthorblockA{ \\
\IEEEauthorrefmark{1} Thales,
\IEEEauthorrefmark{2} Telecom ParisTech \\
}
\IEEEauthorblockA{ \{firstname.name\}@thalesgroup.com \\
\{firstname.name\}@telecom-paristech.fr \\}
}

\maketitle

\begin{abstract}
Current networks more and more rely on virtualized middleboxes to flexibly provide security, protocol optimization, and policy compliance functionalities.
As such, delivering these services requires that the traffic be steered through the desired sequence of virtual appliances.
Current solutions introduce a new logically centralized entity, often called orchestrator, needing to build its own holistic view of the whole network so to decide where to direct the traffic.

\emph{We advocate that such a centralized orchestration is not necessary and that, on the contrary, the same objectives can be achieved by augmenting the network layer routing so to include the notion of service and its chaining.}

In this paper, we support our claim by designing such a system. We also present an implementation and an early evaluation, showing that we can easily steer traffic through available resources. This approach also offers promising features such as incremental deployability, multi-domain service chaining, failure resiliency, and easy maintenance. 
\end{abstract}
\section{Introduction}
\label{sec:intro}

\added[id=A]{Network services used to be built as an ordered set of physically wired hardware appliances that processed traffic for security or optimization purpose. With Network Functions Virtualization (NFV), middleboxes are more and more software-based running on top of virtualization-enabled equipment, thus allowing cost reduction and network  flexibility. Nevertheless, with this new paradigm, new challenges have risen. 
\deleted[id=M]{In particular, since}\added[id=M]{Indeed,} the service function chain\added[id=M]{s are} \deleted[id=M]{is} completely separated from the physical topology, and \deleted[id=M]{because} virtual functions are more ephemeral and dynamic in nature\deleted[id=M]{, s}\added[id=M]{. S}teering traffic through these sparsely located virtual entities, without compromising end-users sessions and Quality of Service (QoS), is \added[id=M]{therefore} a complex challenge.}    



Despite the fact that 
\deleted[id=D]{companies and} 
Internet Service Providers critically rely on middleboxes for security and policy compliance~\cite{Sherry2012}, most of existing NFV \deleted[id=D]{orchestration and} management solutions rely on a logically centralized entity, generally named orchestrator. Such centralized approaches, as they require a holistic view of the whole network to perform service chaining, introduce control reactivity and resiliency (e.g., single point of failure) issues. \added[id=A]{Also, this may be quite costly for operators, since it \added[id=M]{requires}\deleted[id=M]{demands} the deployment of a whole new management and control infrastructure.} In addition, \added[id=A]{the control part, which is meant to modify network configuration so to accommodate the orchestrator decisions,} tend to be poorly interoperable with legacy \added[id=A]{appliances} \deleted[id=D]{networks} and is thus hard to deploy incrementally.  

We believe that centralized orchestration for service function chaining is not necessary. The same functionalities can be provided in a distributed way by directly augmenting the network routing layer. \added[id=A]{In particular, \textit{we argue in this paper that it is possible to leverage on any Interior Gateway Protocol (IGP), anycast addressing, and any service chaining encapsulation, to construct a distributed service-aware distributed control-plane}.}
\added[id=A]{We propose a modular architecture, showing that we do not need complex elements and we remain interoperable with legacy appliances.
We also implemented such architecture, and early evaluation shows that our system successfully steers traffic through the intended service chain, distributing it over different instances according to available resources. }

\added[id=A]{The reminder of this paper is organized as follows. First, we overview related work in Sec.~\ref{sec:relatedwork}. Then, we introduce in Sec.~\ref{sec:concept} the main concept of our proposal: namely~\textit{the service plane topology}. We detail the system architecture in Sec.~\ref{sec:arch} and the implementation in Sec.~\ref{sec:implementation}. Early results supporting our proposal are presented in Sec.~\ref{sec:results}, while Sec.~\ref{sec:agenda} provides an agenda about research worth to be performed with respect to our proposal. Sec.~\ref{sec:conclusion} concludes the paper.}


\section{Related Work}
\label{sec:relatedwork}





So far, NFV frameworks have been built on top of centralized cloud-based management system, which has simplified the use and implementation of resource allocation algorithms \cite{Palkar2015,Herrera2016,Gember2013,Bremler2016}. 
\added[id=M]{For instance, Ghaznavi}~\cite{Ghaznavi2017}\added[id=M]{ proposes a centralized VNF (Virtual Network Functions) splitting and placement algorithm.}
Some solutions, such as Slick \cite{Anwer2015}, go\deleted[id=M]{ even} further proposing a programming language to define, on a central control point, high level policies. Such policies drive the decision being taken and enforced at runtime concerning chaining logic, flow forwarding, VNFs placement. 
Yet, such centralization, while simplifying VNF and path management, comes at the price of increased fragility (e.g., single point of failure, control loop delay, etc.) \added[id=A]{and high} computation load \deleted[id=D]{explosion} on \added[id=A]{the} \deleted[id=D]{this} central point. 

While the Software-Defined Networking (SDN) paradigm helps in simplifying path management, as previously mentioned, it still struggles to achieve traffic steering with fine-grained forwarding rules.
First, \added[id=A]{the size of forwarding state} \deleted[id=D]{the number of forwarding states} is limited by costly memory (TCAM). Second, dealing with forwarding rules installation when there are middleboxes (e.g., NAT -- Network Address Translation -- service) is another challenge to be tackled~\cite{Fayazbakhsh2013,Gember2013,Qazi2013}.
\deleted[id=D]{These centralized approaches are usually complex and tend to introduce some side effects (increased latency, necessity to modify the middlebox etc.).}

Another approach consists in using encapsulation to convey traffic along a service chain.
Recent work at the IETF (Internet Engineering Task Force) proposes Network Service Header (NSH) as a dedicated encapsulation header for service chaining~\cite{RFC8300}.
Segment Routing (SR) encapsulation has also been proposed to handle Service Function Chaining \cite{Abdelsalam2017}. It is based on a source routing model to steer traffic segment to segment.  
Recent work has also made the case for session protocol to build service overlay~\cite{Zave2017}. Even if the approach allows dynamic chaining, it relies on extending/modifying TCP, which, in an ossified Internet, is a hard task \cite{Honda2011}.

Whether or not encapsulation is used, another main challenge in service function chaining is the coordination among flow path and middlebox state. Indeed, sometime VNF instances need to be created or reduced due to fluctuations in flow volume, migrated for resource optimization, or just recovered due to failure.
Some solutions, like OpenNF \cite{Gember2014} and Split/Merge \cite{Rajagopalan2013}, provide an open interface on middleboxes, so to allow a central coordinator to manage both forwarding and state. 
Dysco \cite{Zave2017} proposes to solve these issues by consolidating forwarding and session state into middleboxes. 
Kablan et Al. \cite{Kablan2017} instead try to avoid such state coordination problem by splitting each VNF into a stateless processing part and a consistent back-end data store.

All these works strongly rely on a centralized orchestration or were only used in such context. In the rest of the paper, we make the case for orchestrating service chaining in a distributed manner.
\section{Distributed Orchestration via Network Layer Routing Augmentation}
\label{sec:concept}

While so far service function chaining has relied on a holistic centralized orchestration to steer the traffic through sequences of virtual appliances, we believe that it can be done at the network layer routing in a distributed way. 

Indeed, \textit{any network Interior Gateway Protocol (IGP) can be directly leveraged} to convey the location, the type, and the necessary information associated to a virtual appliance and 
build an augmented network view.


 \begin{figure}[t]
 	\centering
    \subfloat[Network topology.]{%
       \label{fig:IGP-physical}\includegraphics[width=0.95\linewidth]{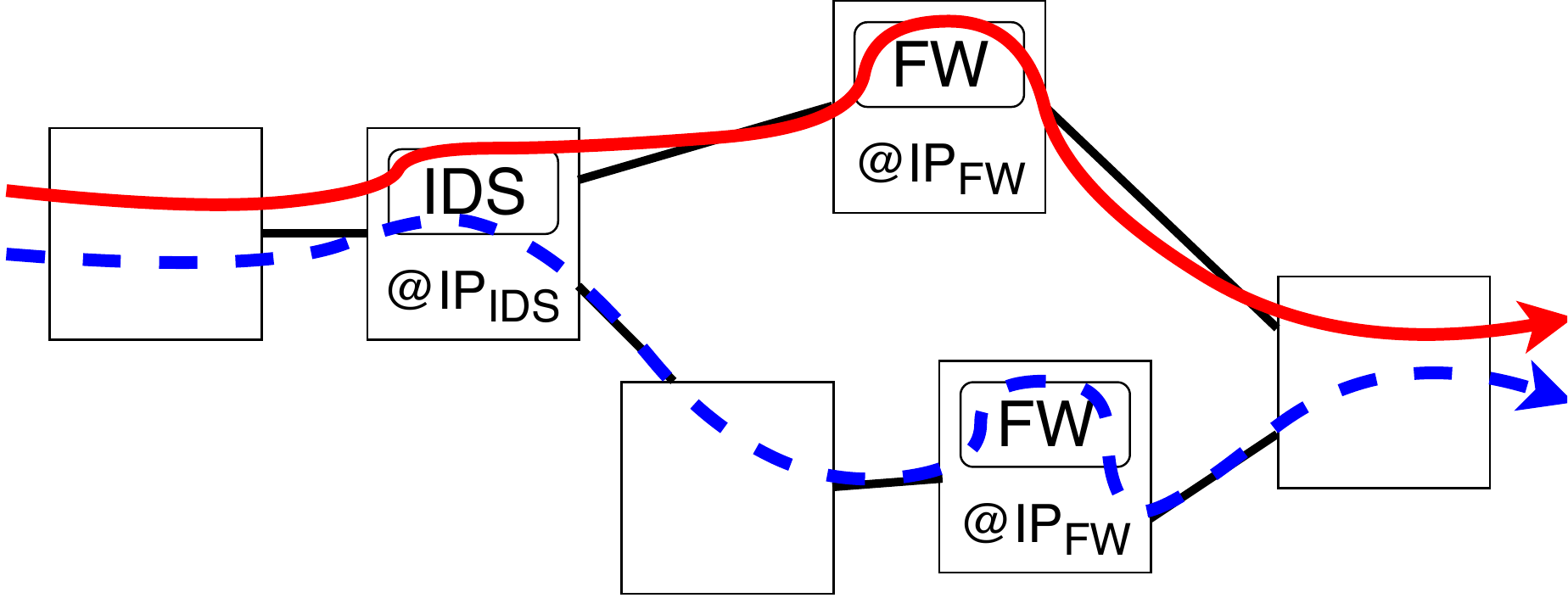}
    }
    
    
    \subfloat[IGP logical view.]{%
        \label{fig:IGP-2-flow}\includegraphics[width=0.95\linewidth]{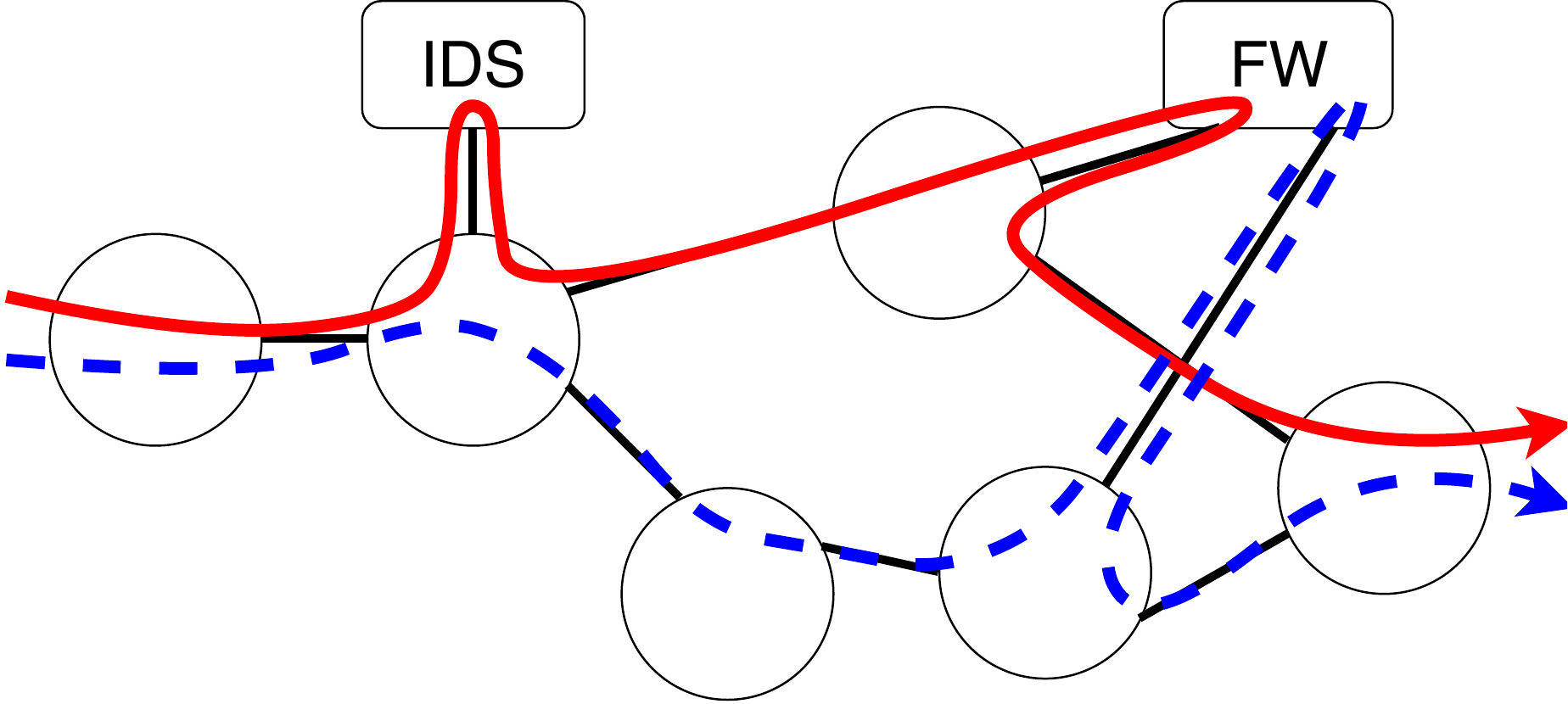}
    }
    \caption{Network topology composed of 6 NFV nodes, with 3 of them hosting VNF instance (\ref{fig:IGP-physical}). The IGP views the two FW instances as a single entity, since they announce the same anycast IP address. A first flow (plain red line) is routed through the IDS and the top FW instance. 
    A second flow (dashed blue line), arriving after the previous one, is then routed through the IDS and the bottom FW instance as the first FW instance is already loaded with the first flow.
    }
    \label{fig:high-level-scenario}
 \end{figure}


Such a view, which we call the \emph{service plane topology}, is formed by two different types of nodes. 
The first type is \emph{NFV nodes}. NFV nodes are physical appliances that run the IGP and host virtual middleboxes (i.e., VNFs). 
NFV nodes can be datacenters, Points of Presence, or routers with VNF hosting capabilities. 
The second type of node, named \textit{VNF node}, corresponds to the VNF instances themselves. NFV nodes can provide different types of service: Deep Packet Inspection (DPI), Firewalling, NAT, stream encoding etc.\deleted{.} These virtual nodes run on top of NFV nodes. Since the NFV nodes that host them run the IGP, \added[id=M]{they can directly use control  packets to share information on their VNF instances. This way,} the VNF nodes are present in the IGP topology \added[id=M]{too}. Consequently, each NFV node has a service plane view\footnote{We use the terms \textit{service plane topology} and \textit{service plane view} interchangeably.} to not only take chaining decisions, but also VNF instantiation, scaling, or deletion decisions.

The main feature of VNF instances is the service they provide. We thus propose to leverage on \textit{anycast addressing} to announce service functions (i.e., \deleted[id=M]{a }VNF instance\added[id=M]{s}) on the network. 
Different VNF instances, that are potentially hosted on different NFV nodes, but that provide the same service, will be announced by the same prefix. This way, each function is mapped to a prefix.
The IGP cost to reach such prefix, announced by each NFV node, can be based on the VNF state, its available capacities, its load, or any other relevant information.\footnote{In Sec.~\ref{sec:agenda}, we discuss more about how to calculate such a metric in a meaningful and rigorous way, so to guarantee loop-free routing convergence.}
A link between two NFV nodes represents a topological distance (network cost), while a link to a VNF node describes some state of the VNF instance (VNF cost). \added[id=A]{Thus, a} routing decision \deleted[id=D]{thus} makes a tradeoff among these metrics and can be designed so as to balance the load, differentiate nodes or chains, etc. \added[id=M]{By applying the routing algorithm associated to the IGP to such a service plane\added[id=M]{ topology}, nodes can populate their routing table, which now becomes service-aware, since it includes routes toward all the available VNF-based services. 
Indeed, VNF instances providing the same service share the same prefix,
hence they can be discriminated by the prefix's attribute(s).} \mb{todo: vérifier les redites sur les prefixes}

\added[id=A]{Figure \ref{fig:high-level-scenario} illustrates with a toy example the approach we propose. Figure \ref{fig:IGP-physical} represents the network topology constituted of NFV nodes. Each VNF instance of a given type is announced on the network with the same anycast address. In particular the two Firewall (FW) instances announce the same prefix: $@IP_{fw}$.  
Flows have to be processed here by a unique chain: $IDS+FW$. The first flow is thus routed through the IDS instance and then through the top FW instance. Indeed, in this example, this VNF instance is at one hop from the NFV node that hosts the IDS instance\deleted[id=M]{ (Figure \ref{fig:IGP-1-flow})}. The NFV nodes that host the used VNFs advertise their neighbors with the new experienced load or any other relevant information. When the second flow arrives, the Firewall instance at the bottom is preferred, resulting in load balancing among the FW instances as well as dynamic path allocation for service function chains (Figure \ref{fig:IGP-2-flow}).  
Notice that in Figure\deleted[id=M]{s \ref{fig:IGP-1-flow}) and} \ref{fig:IGP-2-flow}, since the same prefix is announced but no adjacency is made, the flows that use a link to reach a service function (drawn as boxes) have to use the same link to go out of it. However, note as well that this link is only \emph{virtual}, since it is the representation of the VNF instance in the IGP, but in reality is running directly on a NFV node.}

\deleted[id=M]{\added[id=A]{The proposed system relies on an IGP run on each NFV node to exchange connectivity information and build a topology of the network.
In addition, each NFV node includes on the  IGP control  packets information about the VNF instantiated on it, i.e., the corresponding anycast prefix and their associated attribute(s) (e.g., cost).
Such announces allow to build the \emph{service plane}. 
By applying the routing algorithm associated to the IGP to such a service plane\added[id=M]{ topology}, nodes can populate their routing table, which now becomes service-aware, since it includes routes toward all the available VNF-based services. Indeed, VNF instances providing the same service share the same prefix, hence they can be discriminated by the prefix's attribute(s).}}

Combining this augmented IGP with anycast addressing allows to fully benefit
from what is already done at the network layer routing: network layer information exchange and route computation. Indeed, the NFV nodes can be considered as classic routers that 
compute the next hop(s) for the best path(s), depending on the metrics, to the different \deleted[id=D]{subnetworks} \added[id=A]{prefixes}, which are in our case different network services. 
\deleted[id=D]{It also brings resiliency to the system and ensures compliance with legacy IGP nodes. Finally, the service plane view of the NFV nodes can be directly used to perform either hop-by-hop routing or source routing (i.e., decision at the ingress).}

\added[id=A]{As for any IGP, \textit{high level policies} can be used to control the decision-making at each NFV node. They are common to all the nodes.
Such policies include flow classification rules, to map traffic to the needed service chain. 
High level policies also concern routing decisions since all NFV nodes must share the same routing objectives. Based on the service plane topology, the NFV nodes can use the shortest path algorithm, or any other path computation algorithm, to choose which instance of the next VNF of the chain the flow will go through. Additionally, high level policies can define as well how to compute VNFs' IGP costs, stating which data to use and the function to translate such data in a cost.}

\added[id=A]{To actually drive flows through the service chain they are associated to, we need to leverage on an encapsulation approach. 
In both the hop-by-hop model and the source routing model the encapsulation header should provide the information necessary to steer the flows through the correct sequence of VNFs. 
As such, the header should include \emph{i)} part or all of the service chain identified at the classification step at the ingress of the network and \emph{ii)} the next service step in this chain. For instance, in the example in Figure~\ref{fig:high-level-scenario}, the NFV node that hosts the IDS instance must have a mean to know that a packet belonging to a specific flow has been assigned to the service chain $IDS+FW$, that the next service to apply is $FW$, and which of the $FW$ instances it actually has to go through.}

\section{System Architecture}
\label{sec:arch}

\begin{figure}[t]
\centering
\includegraphics[width=0.83\linewidth]{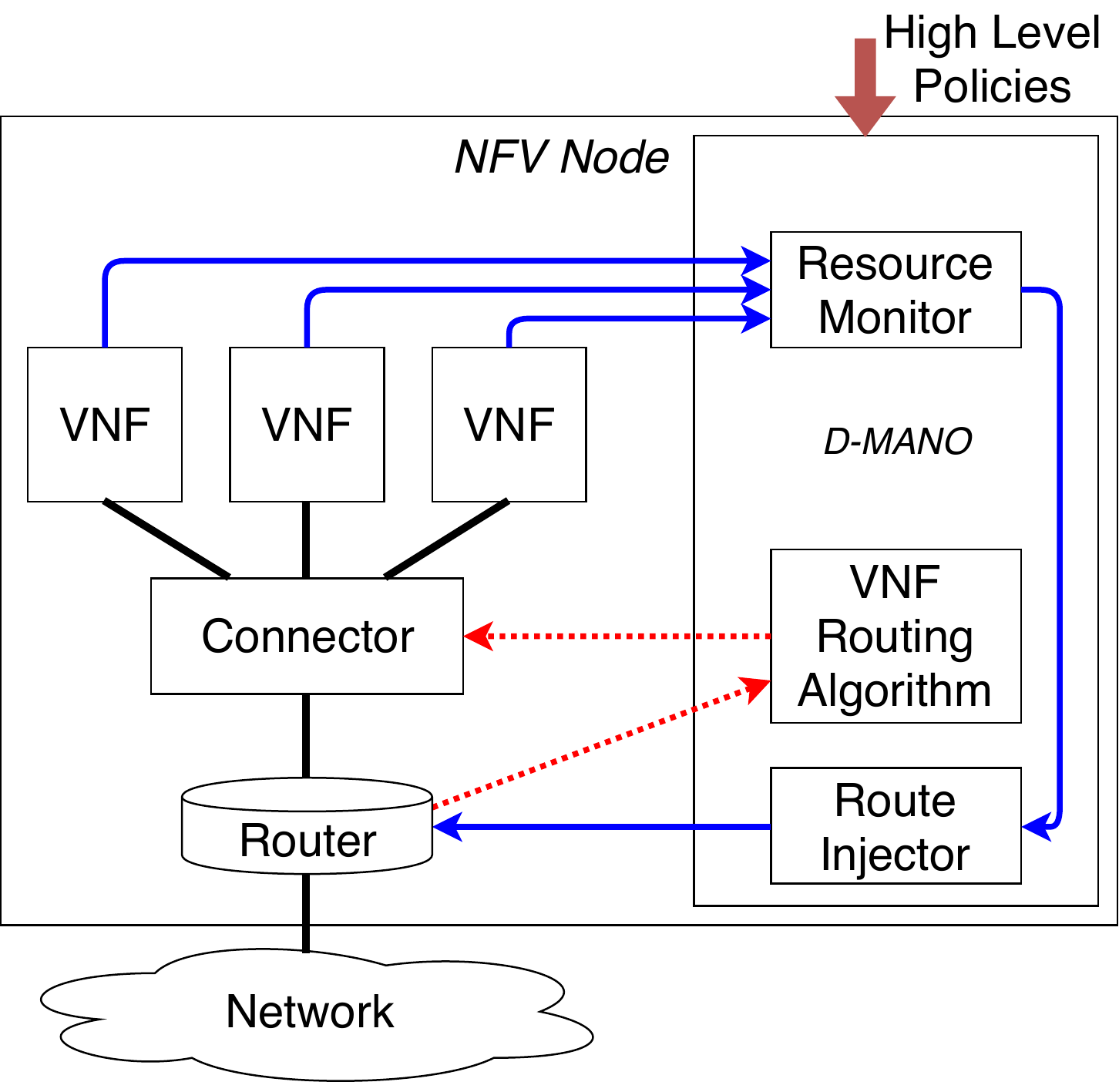}
\caption{NFV Node architecture. Doted arrows illustrate VNF routing control flow. Solid arrows show how VNFs state is monitored, transformed in a cost, which is then injected in the IGP.}
\label{fig:NFV Node archi}
\end{figure}

In this section, we describe the architecture of our system and design its main modules.
A NFV node, as illustrated in Figure \ref{fig:NFV Node archi}, is composed of a \textit{router} providing underlay connectivity, a \textit{connector}, which attaches the router to the different VNF instances, the \textit{VNFs} themselves, providing the services, and a \textit{Distributed MANagement and Orchestration (D-MANO)} component, which allows local autonomous management of the node. 

\textbf{Router:} The router provides both underlay connectivity and participate in the network IGP. 
It exposes a control interface used by the D-MANO to inject or remove VNF anycast prefixes, announcing the services available on the node and the associated costs. This control interface is also used to get the IGP topology to build the service plane topology.

\textbf{Connector:} The connector allows 
dispatching traffic to the VNFs. 
It enforces chaining decisions as follows. 
It forwards incoming packets to the \deleted[id=D]{right} \added[id=A]{intended} VNF instance, based on the encapsulation header. Once the packets have been processed, the VNF forwards them back to the connector, which enforces a forwarding decision toward the next VNF instance location (i.e., its connector) according to the service topology. 
\deleted[id=D]{In an hop-by-hop routing model,}These forwarding decisions are cached in the connector, indexed by a hash computed using \deleted[id=D]{the next service to be reached and} flow-related information. 
\deleted[id=D]{In any case, the connector gets back the packets from the VNF and simply process the encapsulation header. }
The connector also exposes a control interface, used by the D-MANO, to populate the service-aware routing table and the mapping between service function and VNF instance unicast address. This information is used by the connector to enforce chaining decisions for outgoing traffic, and locally balance the load among the VNF instances that provide the same service (same prefix). 



\textbf{VNF:} VNF instances process service flow packets according to the service they provide. 
Once a packet has been processed, the VNF instance updates the chaining encapsulation header to point to the next service.
Each instance is monitored by the D-MANO. 

\textbf{Distributed MANO:} The D-MANO controls and manages the other NFV node's components. It is configured with high level policies, which guide its autonomous orchestration decisions. It has two essential control functions (illustrated in Figure \ref{fig:NFV Node archi}). The first one consists in monitoring VNF instances, deriving from them VNF costs, and then injecting such costs in the IGP, via the router. The second function consists in getting IGP information from the router to build the service plane topology, \deleted[id=D]{computes} \added[id=A]{computing} the service-aware routing table and then \deleted[id=D]{pushes} \added[id=A]{pushing} it in the connector.

\section{Implementation}
\label{sec:implementation}

We started to implement our proposed solution, which we describe in the present section.
Furthermore, we include the technical choices we made for each component of the architecture described in the previous section.

\subsection{System-Level Choices}
\label{subsec:syslevchoices}

\textbf{Encapsulation Header:} 
Our implementation is based on the \textit{Network Service Header (NSH)} protocol to allow steering 
the traffic through the different services~\cite{RFC8300}. 
Even if other encapsulations, such as Segment Routing v6 \deleted[id=D]{protocol} \cite{Abdelsalam2017}, could have been used, our choice is motivated by the fact that NSH is an IETF standard explicitly designed for service chaining and is widely used in many opensource frameworks (e.g., \cite{OPNFV,Opendaylight,ONOS,fdio}). 
In NSH, the Service Path Identifier (SPI) field uniquely identifies a set of abstract service functions (i.e., the Service Function Chain), while the Service Index (SI) points to the next function \added[id=A]{the packet has to be delivered to} \deleted[id=D]{to reach} in the SPI set. NSH also provides extensible metadata fields that we leverage to convey \added[id=A]{the} \deleted[id=D]{a} hash value \deleted[id=D]{. The hash value is} used to consistently identify a flow along its chain\deleted{;}\added{.} \added[id=A]{Such hash value} \deleted[id=D]{and} is computed at the classification step with the 5-tuple of the original packet.

\begin{figure}[t]
\centering
\subfloat[Network topology.]{%
   \label{Fig:weight-cost-a}\includegraphics[width=0.78\linewidth]{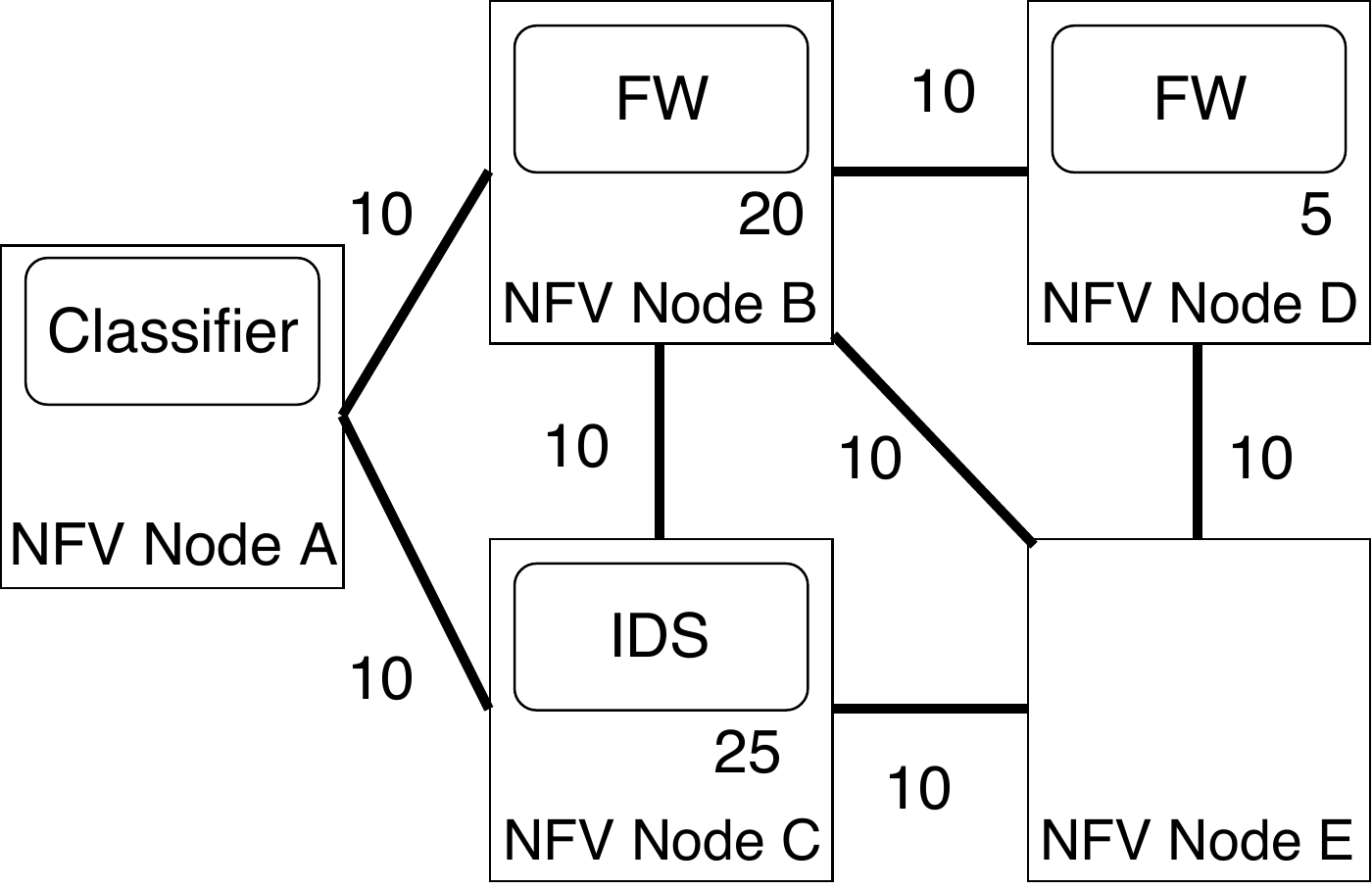}
}

\subfloat[Service plane view at the Node A.]{%
   \label{Fig:weight-cost-b}\includegraphics[width=0.78\linewidth]{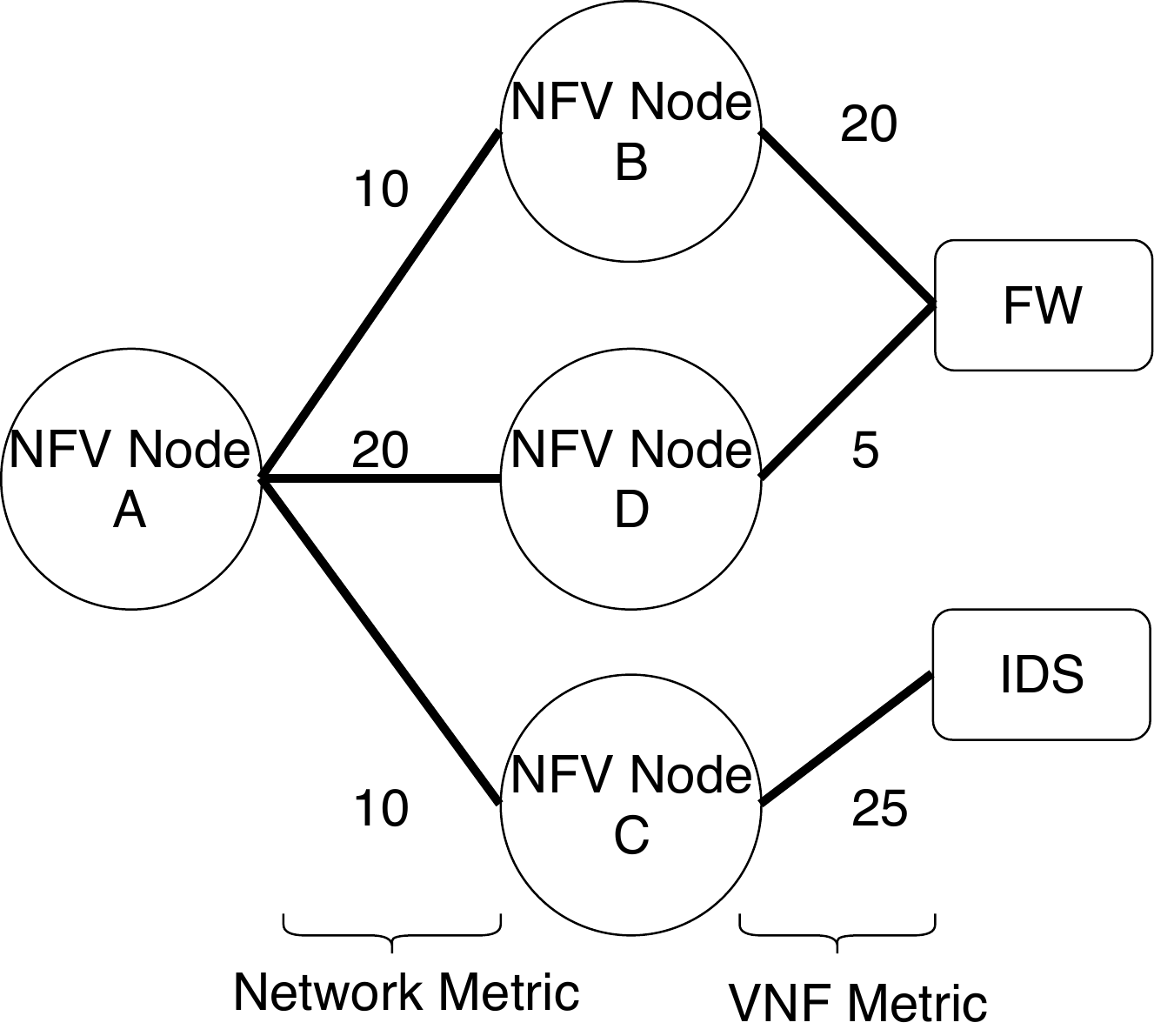}
}
\caption{Each node builds its service plane view (example at Node A on Fig.~(b)) with the Network costs and VNF costs so as to compute the next hop(s).} 
\label{Fig:weight-cost}
\end{figure}

\textbf{IGP:} We build our implementation on top of an \textit{Open Shortest Path First (OSPF)} underlay since this IGP is widely used and easily extensible thanks to opaque Link State Advertisements (LSA). 
Opaque LSAs are leveraged to propagate information about VNF instances and links. Even if flooding opaque LSAs increase control traffic overhead, it does not affect OSPF stability since they do not trigger shortest path algorithm computation. We thus define \textit{VNF opaque LSAs} to convey 3 pieces of data: \emph{i)} the anycast address of a VNF instance, \emph{ii)} the associated VNF cost, and \emph{iii)} the NSH endpoint IP address (i.e., the IP address of the next Connector). In our initial implementation, we choose to use a simple VNF metric: the remaining processing capacity of the VNF instance. The NFV nodes use the provided information to build a graph linking VNFs and NFV nodes, each link being weighted with the associated cost (Fig.~\ref{Fig:weight-cost}). 
Thus, each NFV node is able to build the service plane topology based on the information shared via OSPF. 


\textbf{Service-aware path computation algorithm:} 
We choose to use \textit{Weighted Cost MultiPath (WCMP)}~\cite{Zhou2014} to compute nodes' service-aware routing table. It is particularly suited for our anycast-based approach as it allows balancing 
the traffic based on the VNF cost. 
As illustrated on Figure \ref{Fig:weight-cost}, we use network link costs and VNF costs to weight the paths to a VNF anycast prefix. In this example, \added[id=A]{we show the service topology as seen by node A. It is easy to see that the} cost to reach the $FW$ instance on node B is $30$ and to reach the one on node D is $25$.
\added[id=A]{Such cost is used by WCMP to assign the flows on these instances.} 
Since the VNF cost 
is regularly updated, WCMP adapts to the current load by distributing the traffic on the VNF instances that have the lower load (i.e., lower cost and thus higher WCMP weight). 
In Sec.~\ref{sec:agenda} we discuss more about the metrics.

\subsection{Node-Level Choices}
We build our NFV node using Linux and use network name\-spaces to isolate the components. 


\textbf{Router:} In our implementation, we use \textit{FRRouting} \cite{FRRouting}, an open source IP routing protocol suite, to implement our OSPF router. In particular, 
we use the OSPF API offered by FRRouting to mirror the \deleted[id=D]{router's} Link-State Database (LSDB) in the D-MANO and to inject VNF opaque LSAs. 

\textbf{Connector:}  We implemented the connector logic in \textit{P4}, a language for programming the dataplane~\cite{Bosshart2014}. Our P4 code is run on the simple\_switch target~\cite{simple_switch}. Its runtime CLI is exposed to the D-MANO to configure the switch and populate the WCMP table at runtime. 



\textbf{VNF:} VNFs are implemented as simple processes (using \textit{scapy} \cite{scapy}) parsing incoming packets, decrement their NSH SI field, and forward them back to the connector. The focus of the initial implementation being on the different components of the proposed approach, we purposely choose simplistic VNFs for the time being.
The Python \textit{psutil} library enables us to monitor the resources used by the VNF processes.

\textbf{D-MANO:} The D-MANO has been implemented in Python.
Its main loop runs as follows. First, it polls the resource use of the local VNF instances to build the related costs. The costs are then announced on the network with VNF opaque LSAs. Second, the D-MANO gets the VNF announces from its mirrored LSDB. With these data, it builds a service view (see Fig. \ref{Fig:weight-cost-b}). Based on this topology, it computes WCMP weights and updates them on the connector.

\section{Preliminary Results}
\label{sec:results}

In this section we evaluate a simple scenario to show how we can achieve load balancing on different VNF instances of the same type by using the proposed solution.

We consider a network topology that looks like Figure \ref{Fig:weight-cost-b}, except \added[id=A]{that} we use one single generic service.
Moreover, the link cost between Node A and Node B and between Node A and Node C are set to $160$ and between Node A and Node D it is set to $30$. All the VNF instances have the same initial capacity\added[id=M]{ (i.e., VNF cost)} set to $150$. It is the maximum number of packets per second a VNF instance can process, normalized so to have the same range of values as the link costs. We use Mininet to emulate this topology \cite{Lantz2010}.

Traffic has to be steered through \added[id=M]{one of the VNF instances}\deleted[id=M]{a VNF} and then toward the egress. We generate constant bit-rate flows on a source connected to Node A. Each flow lasts $50$ seconds and consumes $2$ of processing units at the VNFs. The arrival rate is of two flows per second.
Our scenario evolves in \added[id=M]{2}\deleted[id=M]{two} phases. \textit{Phase 1:} only the VNFs on Node B and Node C are running. \textit{Phase 2:} After 150 seconds, a third VNF is instantiated on at Node D, which leads\added[id=M]{ to} a redistribution of the traffic, since there are more VNF processing capacit\added[id=M]{ies}\deleted[id=M]{y} now available in the service topology.  

\begin{figure}
\centering
    \includegraphics[width=1\linewidth]{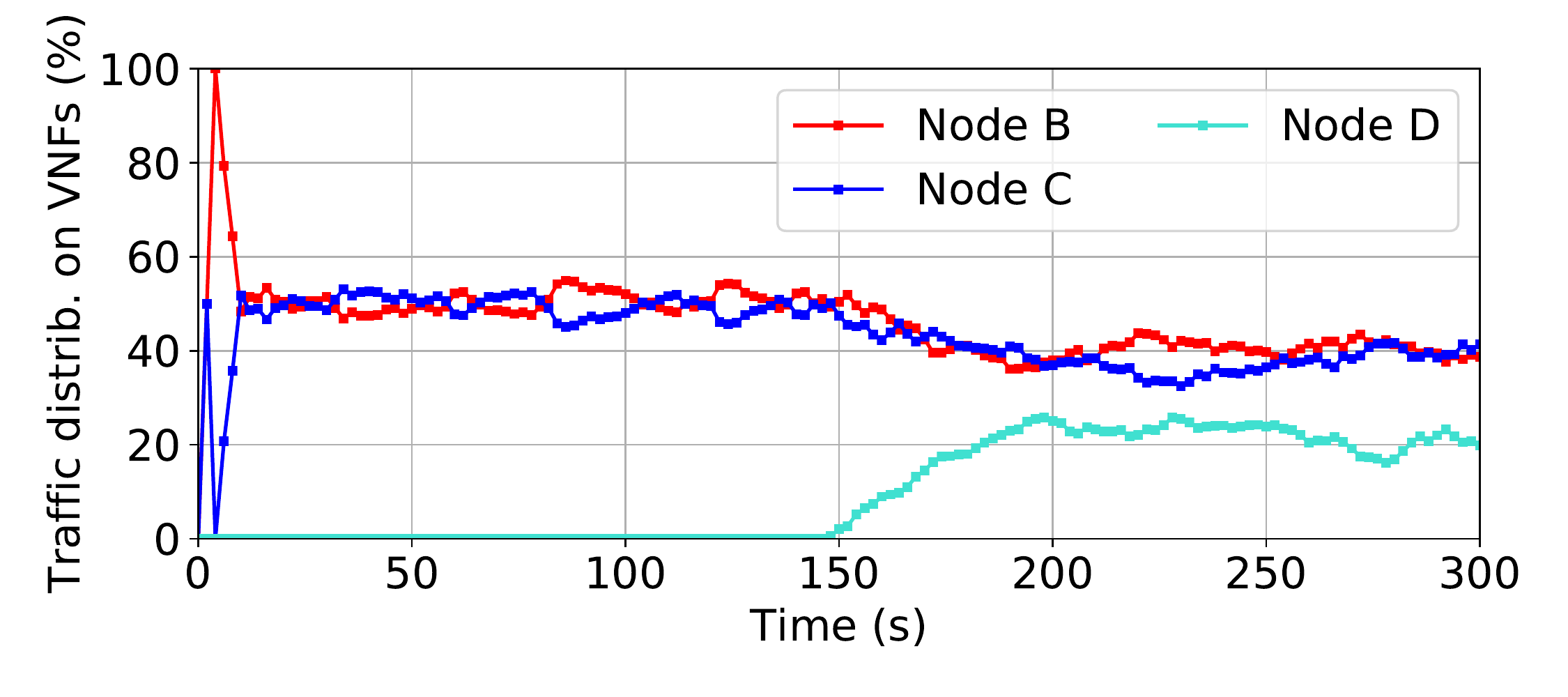}
\caption{Traffic distribution over time on the VNF instances. During the first 150s only two VNFs are running. At $t=150s,$ a third VNF is instantiated.}
\label{Fig:lb_static}
\end{figure}

Figure \ref{Fig:lb_static} presents the traffic distribution over time on the VNF instances.
Since each flow lasts 50s, during the first 50s of the experiment, the system load rises until it reaches its steady state. Note that, in this example, a measure of each VNF load is measured and advertised every 2s. We can see that, during the first phase, each VNF instance receives in average the same amount of traffic. Indeed, they do have the same network cost from the ingress point of view and the same initial VNF cost. 
Once Phase 2 starts, after the 50 seconds of transition, which lasts between $t=150s$ and $t=200s$, a new steady state is reached. Now the VNFs on Node B and C, each process $40\%$ of the traffic, while the VNF on Node D roughly processes $20\%$. This distribution of traffic corresponds to the WCMP weights that consider links' cost and VNFs' cost.

\begin{figure}
\subfloat[Phase 1 (50-150s).]{%
    \label{fig:distrib1}\includegraphics[width=0.5\linewidth]{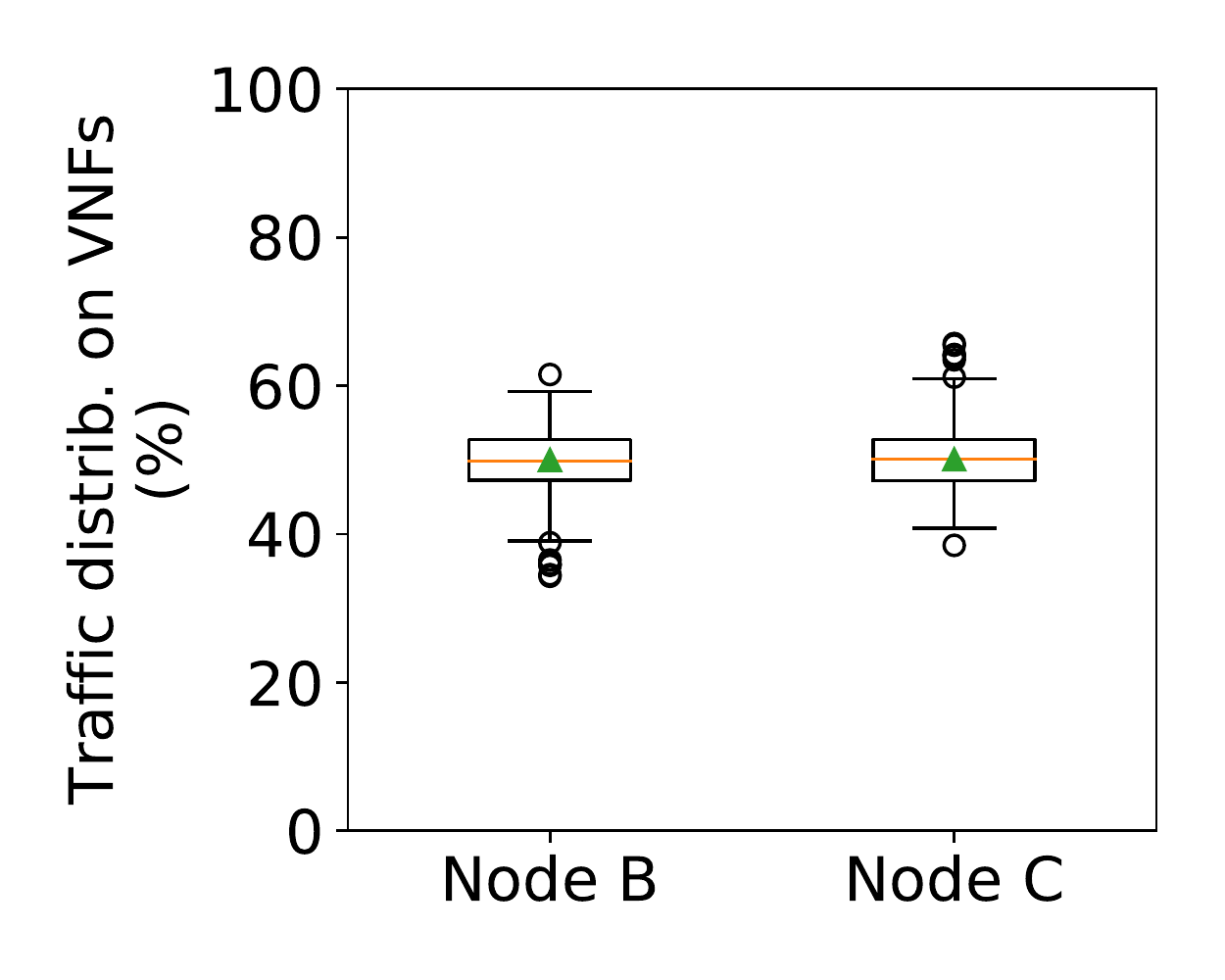}
}
\subfloat[Phase 2 (200-300s).]{%
    \label{fig:distrib2}\includegraphics[width=0.5\linewidth]{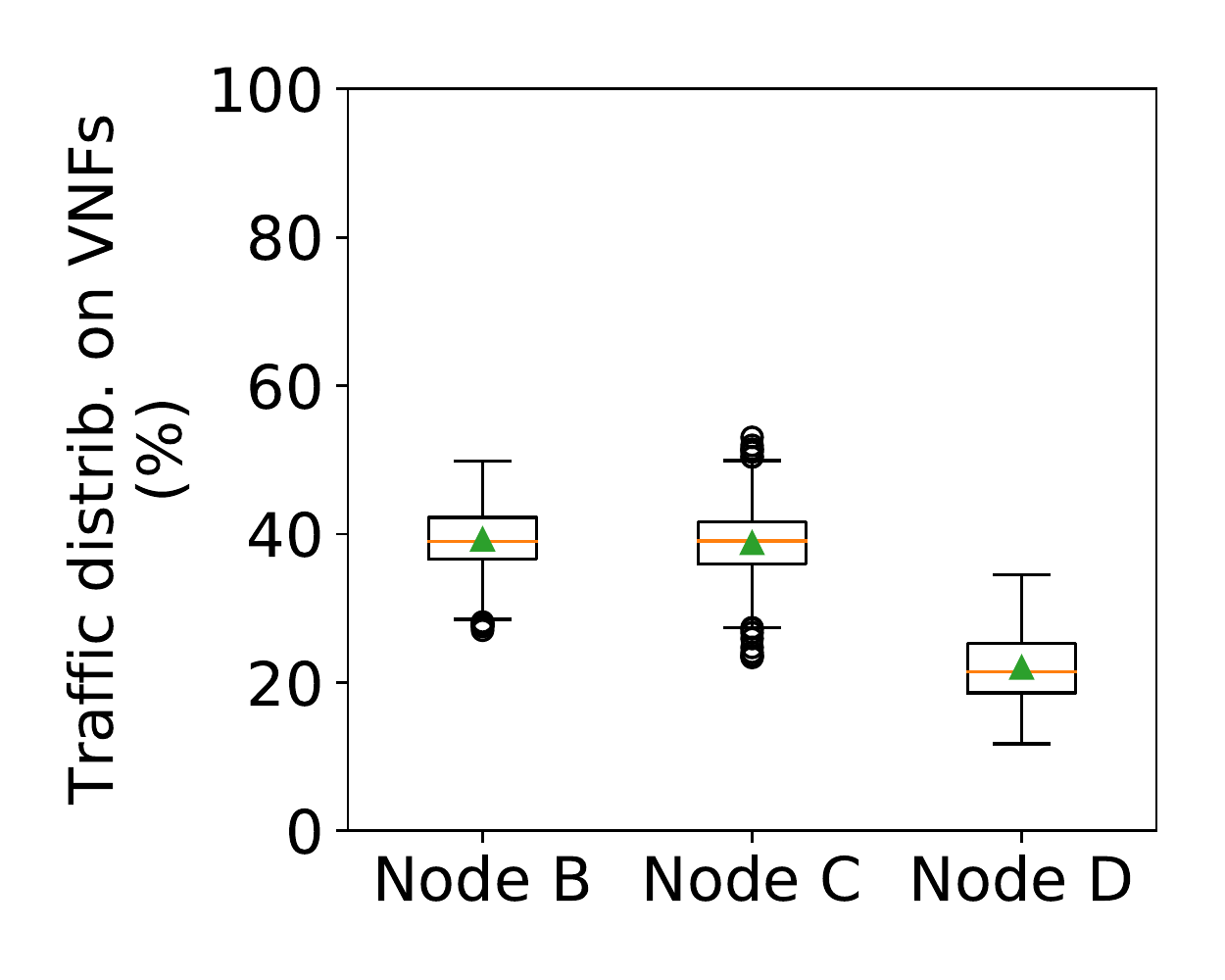}
}
\caption{Boxplot of the traffic distribution on the VNF instances during the two phases.}
\label{fig:2phases}
\end{figure}

Figure \ref{fig:2phases} presents the mean traffic distribution on the instances for the steady state of the two phases of the scenario. They result from 20 runs of the experiment. We can observe that \added[id=A]{our solution is able to balance the load among the available VNFs.} \deleted[id=D]{the system well balances the traffic:}\added[id=M]{T}\deleted[id=M]{t}he mean and median loads are centered \deleted{around}\added{on} the values we can compute from WCMP: $50/50\%$ in Phase 1 and $40/40/20\%$ in Phase 2. In addition, $50\%$ of the loads are less than $3$ points from the median value, while the max and min values are at most $10$ points from it. \added[id=A]{Such limited variation shows that the system remains quite stable}. 

\section{Research Agenda}
\label{sec:agenda}


Our preliminary results illustrate how service chaining can indeed be achieved by augmenting the network layer routing and applying high level policies.
However, while opening interesting perspectives, it opens as well a number of questions. We overview them in this section.


\textbf{Traffic Engineering Constraints:}
Forwarding traffic \added[id=A]{in} \deleted[id=D]{through} service function chains, fulfilling both Service Level Agreements and cost minimization, is a hard task. Some long-lived flows require QoS guarantees (e.g., small delay for VoIP), while short ones may suffer from an initial latency in path computation (e.g., DNS query). 
We believe that best effort traffic and short lived flows are best handled by precomputed hop-by-hop routing decisions. Conversely, traffic \added[id=M]{requiring}\deleted[id=M]{needing} resource reservation would be best served using the source routing paradigm. 
Our service plane topology provides support for both approaches, and enables enforcing high-level policies. However, such hybrid scenarios and related tradeoff need further investigation. 


\textbf{VNF Metrics:} In our approach, service-aware routing involves two different types of entity: namely network links and VNF instances. While assigning a cost to a link is \added[id=A]{straightforward and normal operation} \deleted[id=D]{classic} (based on bandwidth, latency etc.), evaluating the \emph{cost} of a VNF instance is an open research area. On the one hand, such a cost may be based on a plethora of VNF state parameters~\cite{Cao2015,Naik2016}. On the other hand, the metric computation needs to be in the same order of magnitude of the links' metric, and, more importantly, it has to be additive, so to guarantee loop-free convergence even when taking into account multiple constraints~\cite{Wang1996,Jaffe1984}.

\textbf{VNF/Resource Management:} 
Resource allocation is already a hard problem when using a centralized approach~\cite{Herrera2016}. 
Even if a distributed approach, like ours, improve\added{s} the architecture resiliency and scalability, it adds coordination to the problem.  
In a distributed environment, each NFV node needs to take autonomous VNF provisioning decisions based on the exchanged information. Defining the needed information, their granularity,\added[id=M]{ their update frequency,} and the range of actions that each NFV node can take according to global resources availability remains to be explored. However, there is a great potential to use Machine Learning solutions that would make the network completely autonomous.


\textbf{Service Modification:} During a flow lifetime, the service chain it is associated to may be modified for numbers of reasons
(e.g., a suspicious flow is redirected to a DPI). 
Such a service change implies a traffic redirection. 
In \added[id=A]{the source routing model,} \deleted[id=D]{source routing,} this modification could be easily handled since per flow state \deleted{are}\added{is} concentrated at the edge. 
Conversely, in the hop-by-hop \added[id=A]{model,} \deleted[id=D]{routing,} the chaining protocol should coordinate \added[id=M]{NFV nodes' state}\deleted[id=M]{ NFV node states} in order to modify the flow path (e.g., use Operations, Administration, and Maintenance for signaling). 
Existing work \cite{Zave2017} identified challenges to keep end user sessions alive during reconfigurations\deleted[id=D]{(removing proxy, data loss etc.)}. Coordination between VNF session state and routing decisions remain\added[id=M]{s} a challenging question to be explored. 

\textbf{Maintenance and Failure:} To provide carrier grade network services, 
the impact of VNF unavailability on existing traffic should be minimized.
With our approach maintenance can be easily handled through any existing loop-free graceful shutdown approach~\cite{Francois2007}.
Furthermore, some VNF state migration use-cases can be dealt locally with on NFV nodes~\cite{Rajagopalan2013,Kablan2017,Woo2018}. 
However, VNF migration to a remote NFV node is more challenging. 
Indeed, NFV state migration has to be coordinated with service topology update. Certainly existing fail-over mechanisms and \deleted{make-before-brake} \added{make-before-break} approaches can be considered, yet, the design of such mechanism is an open research area.

\textbf{Security:} Distributing the service chaining decision raises some inherent security questions. To state if a VNF announce is valid or not, NFV nodes should trust each other\deleted{s}. Trustworthiness can be solved by key distribution and initial authentication. Moreover, since the chaining protocol may convey sensitive information in its header or metadata, it may be useful to use encryption between authenticated NFV nodes. For instance, we could use IPsec~\cite{rfc4301} as a transport encapsulation between NFV nodes. In general, since we are augmenting the network routing layer, without revolutionizing it, there is quite a number of existing security solutions that can be considered to provide security in the service topology. 


\textbf{Multi-domain SFC:} Even if multi-domain SFC would open new business opportunities, service providers are reluctant to share information related with their network. We believe that a distributed design can ease multi-domain orchestration. Defining an IGP routing logic to provide distributed SFC decisions is a first step for the design of multi-domain services. Indeed, the next step to be investigated is the use of inter-domain routing, based on BGP~\cite{RFC4271} to provide chaining among different administrative entities, for instance based on the use of communities~\cite{RFC1997}. 


\section{Conclusion}
\label{sec:conclusion}

In this paper, we have made the case for orchestrating service chaining in a distributed manner. We proposed to augment the network layer routing by using anycast addressing for VNF so to build what we call the service topology, allowing embedding service chaining into routing.
We designed an architecture based on this concept and implemented a first \added[id=A]{prototype.} \deleted[id=D]{version of it.}
Early evaluation performed with our implementation shows that flows can be successfully driven through the chain of services according to available resources.
Our approach sets itself apart from previous work, and as such it still needs to be \deleted{throughly}\added{thoroughly} investigated. To this end we provide a research agenda highlighting the different aspects that need to be tackled. However, what comes out as well is quite promising and opens interesting perspectives.



\bibliographystyle{abbrv} 
\begin{small}
\bibliography{Adrien-Bibliography}
\end{small}

\end{document}